\documentclass[11pt]{article}
\usepackage{graphicx,subfigure,amsmath,amsthm,mathrsfs,amssymb}
\newcommand{\ee}{\end{equation}}
\newcommand{\word}[1]{\,\,\mbox{#1}\,\,}
\newcommand{\reff}[1]{(\ref{#1})}
\newcommand{\beq}{\begin{equation}}
\newcommand{\eeq}[1]{\label{#1}\end{equation}}
\newcommand{\beg}{\begin{equation*}}
\newcommand{\eeg}{\end{equation*}}

\newcommand{\fivequad}{\qquad\qquad\qquad\qquad\qquad}
\newcommand{\eq}{\!=\!}
\newcommand{\p}{\!+\!}
\newcommand{\m}{\!-\!}
\newcommand{\too}{\!\to\!}

\newcommand{\bsplit}{\begin{split}}
\newcommand{\esplit}{\end{split}}

\usepackage[usenames]{color}
\usepackage{ulem}
\usepackage[square,comma,sort&compress]{natbib}

\begin{document}
\def\theequation{\arabic{section}.\arabic{equation}}
\begin{titlepage}
\title{\vspace{-2em} $3D$ scalar model as a $4D$ perfect conductor limit:
\\
dimensional reduction and variational boundary conditions}
\author{$^{1,2}$Ariel Edery\thanks{Email: aedery@ubishops.ca}\,\,,\quad $^{2,3}$Noah Graham\thanks{Email:
ngraham@middlebury.edu}\,\,,\quad $^{4,1}$Ilana
MacDonald \thanks{Email: macdonald@astro.utoronto.ca}\\\\$^1$ {\small\it Physics Department, Bishop's University}\\
{\small\it 2600 College Street, Sherbrooke, Qu\'{e}bec, Canada
J1M~0C8}\vspace{0.5em} \\ $^2$\thanks{work partly completed at KITP,
Santa Barbara} {\small\it Kavli Institute for Theoretical Physics,
 University of California}\\{\small\it Kohn Hall, Santa Barbara,
CA 93106 USA } \vspace{0.5em} \\$^3${\small\it Department of
Physics, Middlebury College, Middlebury, VT 05753} \vspace{0.5em}
\\$^4$\thanks{present address} {\small\it Department of Astronomy
and Astrophysics, University of Toronto}\\{\small\it 50 St. George
Street, Toronto, Ontario Canada M5S 3H4
 }}
\date{} \maketitle
\vspace*{-1truecm}
\begin{abstract}
Under dimensional reduction, a system in $D$ spacetime dimensions
will not necessarily yield its $D\m1$-dimensional analog version.
Among other things, this result will depend on the boundary
conditions and the dimension $D$ of the system. We investigate this
question for scalar and abelian gauge fields under boundary
conditions that obey the symmetries of the action. We apply our
findings to the Casimir piston, an ideal system for detecting
boundary effects. Our investigation is not limited to extra
dimensions and we show that the original piston scenario proposed in
$2004$, a toy model involving a scalar field in $3D$ ($2\p1$)
dimensions, can be obtained via dimensional reduction from a more
realistic $4D$ electromagnetic (EM) system. We show that for perfect
conductor conditions, a $D$-dimensional EM field reduces to a $D\m1$
scalar field and not its lower-dimensional version. For Dirichlet
boundary conditions, no theory is recovered under dimensional
reduction and the Casimir pressure goes to zero in any dimension.
This ``zero Dirichlet" result is useful for understanding the EM
case. We then identify two special systems where the
lower-dimensional version is recovered in any dimension: systems
with perfect magnetic conductor (PMC) and Neumann boundary
conditions. We show that these two boundary conditions can be
obtained from a variational procedure in which the action vanishes
outside the bounded region. The fields are free to vary on the
surface and have zero modes, which survive after dimensional
reduction.
\end{abstract}
\setcounter{page}{1}
\end{titlepage}

\def\theequation{\arabic{section}.\arabic{equation}}

\section{Introduction}

In many applications one considers the result of dimensional
reduction, in which one dimension of a field theory is made small
(or reduced to zero) with suitable boundary conditions imposed along
this dimension. In this paper, the dimension we reduce is not
``curled up", but instead is taken along an interval with boundary
conditions that respect the symmetries of the original action
(typically Lorentz and gauge invariance). This process of
dimensional reduction is not limited to extra dimensions; it applies
equally well to $4D$ $(3\p1)$ systems with material boundaries, as
long as they are idealized so that the symmetries of the action are
obeyed. The original action is decomposed into massless and massive
sectors of one lower dimension after the boundary conditions are
included and one dimension is ``integrated out''. The effective
action after dimensional reduction is then obtained. We apply these
results to a particular physical system, the Casimir piston, and
show that the original Casimir piston scenario introduced in 2004
\cite{Cavalcanti} for a $3D$ scalar field obeying Dirichlet boundary
conditions can be obtained via dimensional reduction from a $4D$
electromagnetic (EM) system obeying perfect conductor conditions.
Simply put, a toy model involving a $3D$ scalar field emerges from a
more realistic $4D$ EM system.

A question of general interest is whether the lower-dimensional
version of a system is recovered under dimensional reduction. For
example, consider a massless scalar field in $d\p1$ dimensions with
one dimension compactified to a circle of radius $R$. Sending $R\to
0$ yields its lower-dimensional version, a $d$-dimensional scalar
field, because the Fourier decomposition of the field includes an
$n\eq0$ mode, which yields exactly the $d$-dimensional scalar field.
The non-zero modes become infinitely massive as $R\to 0$ and can be
ignored. This scenario, however, does not apply here, since the
dimension we reduce is not curled up. The boundary conditions we
consider are perfect magnetic conductor (PMC) and perfect electric
conductor (PEC) conditions for abelian gauge fields and Dirichlet or
Neumann boundary conditions for scalar fields. In only two of these
four cases does one recover the lower-dimensional version of the
field under dimensional reduction.

We begin our study with massless abelian gauge (EM) fields obeying
perfect electric conductor (PEC) conditions. We show that this
system does not reduce to its lower-dimensional version under
dimensional reduction except in $4D$. We then apply our results to
the Casimir piston. Besides being an ideal system for detecting
purely boundary effects, work carried out in the last four years has
provided formulas for the $3\p1$ Dirichlet piston \cite{Edery},
$3\p1$ EM piston \cite{Kardar,Marachevsky} and higher-dimensional
non-compactified scenarios \cite{Ariel-Ilana, Ariel-Valery-PMC,
Ilana-Thesis}. There has also been a large amount of recent work in
this field \cite{Saharian,Kirsten,Cheng,Teo,Barton} (see also the
introduction to \cite{Ariel-Valery-PMC} for some historical
details). Dimensional reduction in pistons for scalar fields in
Kaluza-Klein scenarios has recently been discussed in
\cite{Kirsten}.

We show that Cavalcanti's original piston scenario
\cite{Cavalcanti}, a toy model involving a $3D$ scalar field, can be
obtained via dimensional reduction from a realistic $4D$
electromagnetic system. We then investigate scalar fields under
Dirichlet boundary conditions. Under dimensional reduction, no
theory is recovered and the Dirichlet Casimir piston yields zero
Casimir force. We show that this ``zero Dirichlet" result is useful
for understanding other systems like the PEC system. Perfect
magnetic conductor (PMC) conditions are dual to PEC conditions and
obey the same symmetries, namely Lorentz and gauge invariance. PMC
conditions for EM fields and Neumann conditions for scalar fields
have the same special property under dimensional reduction: they
yield their lower-dimensional versions in any dimension. We explain
this phenomenon by showing that they are obtained through a
variational procedure in which the action vanishes outside the
bounded region but with the field not fixed at the boundary surface.
In both cases, the fields have zero modes, which survive after
dimensional reduction.

\section{Dimensional reduction of a PEC system: $4D$ EM to $3D$ scalar}

Perfect electric conductor (PEC) boundary conditions can be
generalized to any dimension. In a $d\p1$ dimensional spacetime they
are given by \beq
\eta^{\mu}F^{\ast}_{\mu\,\alpha_1\,\alpha_2\,\alpha_{d -2}}=0
\eeq{PEC} where $\eta^{\mu}$ is a spacelike vector normal to the
bounded hypersurface. $F^*$ is the dual to the field strength
$F_{\mu\nu}=\partial_{\mu} A_{\nu}-\partial_{\nu}A_{\mu}$ and is
defined by \beq F^{\ast}_{\alpha_1\,\alpha_2\,\ldots\,\alpha_{d-1}}
\equiv\varepsilon_{\alpha_1\,\alpha_2\,\ldots\,\alpha_{d-1}\,\mu\,\nu}\,F^{\mu\nu}\eeq{dual}
where $\varepsilon_{\alpha_1\,\alpha_2\,\alpha_{d-1}\mu\nu}$ is the
$d\p1$ dimensional Levi-Civita tensor. The PEC conditions \reff{PEC}
are Lorentz and gauge invariant and hence preserve the symmetries of
the higher-dimensional Maxwell action. In $3\p1$ dimensions, they
yield the familiar boundary conditions at the surface of a perfect
conductor: ${\bf n}\times {\bf E}=0$ and ${\bf n}\cdot{\bf B}=0$,
where ${\bf E}$ and ${\bf B}$ are the electric and magnetic fields
respectively and ${\bf n}$ is the vector normal to the surface.

Consider two parallel hyperplanes situated at $x^d\eq 0$ and $x^d\eq
L$ with normal vector $\eta^{\mu}$ in the $x^d$ direction. The
following mode decomposition for the gauge fields satisfy the PEC
condition \reff{PEC} at the two planes: \beq\begin{aligned}
A_{\mu}(x^{\mu},x^d)&= \sum_{n=1}^{\infty}A_{\mu}^{(n)}(x^{\mu})
\,\sin\,(n\,\pi\,x^d/L)\quad\quad\mu=0,1,\ldots,d\m1\\A_{d}(x^{\mu},x^d)&=\sum_{n=0}^{\infty}A_{d}^{(n)}(x^{\mu})
\,\cos\,(n\,\pi\,x^d/L)\\&=A_d^{(0)}(x^{\mu})+
\sum_{n=1}^{\infty}A_{d}^{(n)}(x^{\mu})
\,\cos\,(n\,\pi\,x^d/L)\,.\end{aligned}\eeq{Decomp}

One can to go to axial gauge $A_d\eq0$ but it is more convenient to
go to ``almost" axial gauge \cite{Raman} where
$A_{d}=A_d^{(0)}$\footnote{One cannot eliminate the zero mode field
$A_d^{(0)}$ via a gauge transformation. If one uses axial gauge
$A_d\eq0$ instead of ``almost" axial gauge, then $A_d^{(0)}$ appears
in the new $A_{\mu}$ and one obtains the same action as
\reff{action2} though it is slightly longer to derive. See appendix
A.}. This can be achieved with the gauge function \beq\Lambda=
\sum_{n=1}^{\infty}-\dfrac{L}{n\,\pi} \,\,A_{d}^{(n)}(x^{\mu})
\,\sin\,(n\,\pi\,x^d/L)\,.\eeq{Lambda} Note that after the gauge
transformation, $A_{\mu}$ retains the same form. The mode
decomposition in this gauge is given by \beq\begin{aligned}
A_{\mu}(x^{\mu},x^d)&= \sum_{n=1}^{\infty}A_{\mu}^{(n)}(x^{\mu})
\,\sin\,(n\,\pi\,x^d/L)\\A_{d}&=A_d^{(0)}(x^{\mu})\equiv\phi(x^{\mu})\end{aligned}\eeq{Amu}
where $\phi(x^\mu)$ represents a scalar field. Our metric signature
is $(+,-,-,...,-)$ so that $A^d=-\phi$.

The generalized Maxwell action in $d\p1$ dimensions is given
by \beq\begin{aligned} S&=\int -\dfrac{1}{4} \,F_{MN}F^{MN}
\,d^{\,d+1}x &M,N=0,1,\ldots,d&\\
&=\int -\dfrac{1}{4} \,F_{\mu\nu}F^{\mu\nu} \,d^{\,d+1}x +\int
-\dfrac{1}{2} \,F_{\mu\,d}F^{\mu\,d} \,d^{\,d+1}x
\qquad&\mu,\nu=0,1,\ldots,d-1&.
\end{aligned}\eeq{action}
After substituting the mode decomposition \reff{Amu} into
\reff{action} and integrating over $x^d$ from $0$ to $L$, we obtain
the following action: \beq\begin{aligned} S&= \dfrac{L}{2}\int
\,\partial_{\mu}\phi\,\partial^{\,\mu}\phi \,\,d^{\,d}x+
\dfrac{L}{2}\int \sum_{n=1}^{\infty}\left\{-\dfrac{1}{4}
\,F_{\mu\nu}^{(n)}F^{{\mu\nu}^{(n)}} + \dfrac{1}{2}
\dfrac{n^2\pi^2}{L^2}\,A_{\mu}^{(n)}A^{{\mu}^{(n)}}\right\}d^{\,d}x
\end{aligned}\eeq{action2}
where $F_{\mu\nu}^{(n)}\equiv \partial_{\mu}
A_{\nu}^{(n)}-\partial_{\nu}A_{\mu}^{(n)}$. The original
$d\p1$-dimensional Maxwell action has decomposed into a massless
$d$-dimensional massless scalar field $\phi(x^{\mu})$ and an
infinite tower of $d$-dimensional massive spin 1 fields
$A_{\mu}^{(n)}$ of mass $m_n=n\pi/L$. Under dimensional reduction,
i.e. as $L\to 0$, the spin 1 modes become infinitely massive and the
theory reduces to a $d$-dimensional massless scalar field.
Therefore, under PEC conditions, the lower-dimensional version of
the original system, a $d$-dimensional EM field, is not recovered
after dimensional reduction. There is, however, one exception. A
$d$-dimensional EM field has $d\m2$ degrees of freedom and therefore
has one degree of freedom when $d\eq3$. In other words, a $3D$ EM
field and a $3D$ scalar field both have one degree of freedom and
the two can be thought to be equivalent (as long as the boundary
conditions match). For PEC conditions there is no other dimension
where the lower-dimensional version is recovered.

\begin{figure}[ht!]
\begin{center}
\subfigure[Piston geometry in two spatial dimensions. The length $s$
is taken to be infinite, $a$ is the plate separation and $b$ is the
length of the second side. The $3D$ Dirichlet piston is a scalar
field obeying Dirichlet boundary conditions on all
sides.]{\label{3D}
\includegraphics[scale=0.50]{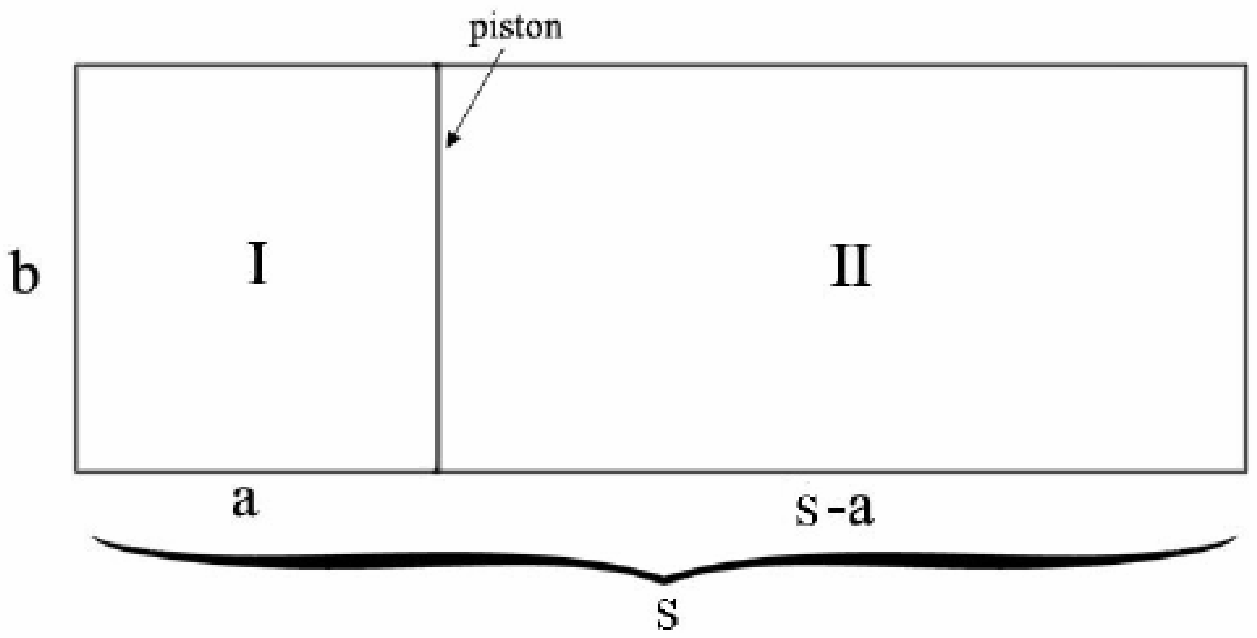}}\hspace{1em}
\subfigure[Piston geometry in three spatial dimensions. The plate separation is $a$ and $s$
is taken to be infinite. The $4D$
PEC piston is an EM field
obeying perfectly conducting conditions on all walls.  In the
limit as one of the sides (b or c) goes to zero, we recover the
$3D$ Dirichlet piston.]{\label{4D}
\includegraphics[scale=0.75]{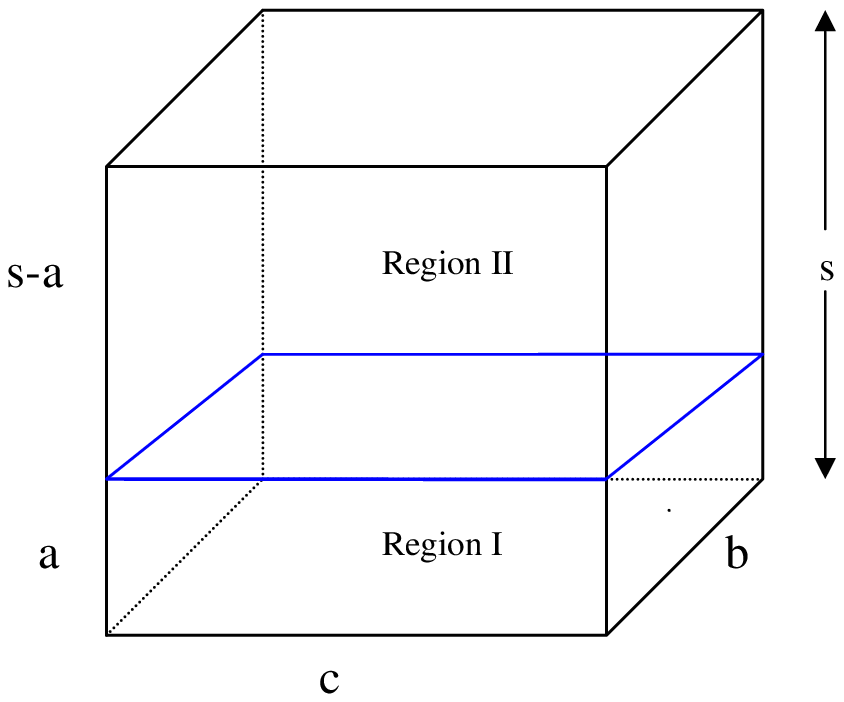}}
\caption{piston geometry}
\end{center}
\end{figure}

In 2004, a piston geometry was introduced for Casimir calculations
\cite{Cavalcanti}. The piston separates two regions, each of which
contributes to the Casimir force on the piston. The original
scenario was a toy model involving a $3D$ scalar field obeying
Dirichlet conditions in a rectangular cavity (see Fig.\,\ref{3D}).
Later, a more realistic system, the $4D$ PEC piston (see
Fig.\,\ref{4D}) was solved exactly \cite{Kardar, Marachevsky,
Ariel-Valery-PMC, Ilana-Thesis}. We now show that these two systems
are related: the original toy model, the $3D$ Dirichlet Casimir
piston, can be viewed as a limiting case of the more realistic $4D$
PEC Casimir piston as one dimension is reduced. The Casimir force on
a piston for PEC conditions in $4D$ with plate separation $a$ and
sides $b$ and $c$ can be expressed in many equivalent but different
forms. The form found in \cite{Ariel-Valery-PMC, Ilana-Thesis} is
the most convenient for our purposes: \beq\begin{split}
 F_{\!P\!E\!C}&=\dfrac{1}{2\,c}\,\sum_{n=1}^{\infty}\sum_{\ell=1}^{\infty}\dfrac{n}{\ell}
\dfrac{\partial\,}{\partial\,a}K_1\big(\frac{2\,\pi\,n\,\ell\,a}{c}\big)\\&\qquad\qquad
+\,\dfrac{\partial\,}{\partial\,a}
\Bigg\{\dfrac{a\,c}{2}\sum_{n=1}^{\infty}\,\sum_{\ell_1=1}^{\infty}\sum_{\substack{\ell_2=-\infty\\}}^{\infty}
\Big(\dfrac{n}{b}\Big)^{3/2}\,\dfrac{\,K_{\frac{3}{2}}
\big(\,\frac{2\pi\,n}{b}\,\sqrt{(\ell_1\,a)^2+ (\ell_2\,c)^2}
\,\big)}{\left[(\ell_1\,a)^2
+(\ell_2\,c)^2\right]^{\tfrac{3}{4}}}\Bigg\}\,.
\end{split}
\eeq{FPEC} The above formula is invariant under the exchange of $b$
and $c$ \cite{Ariel-Valery-PMC}. Without loss of generality, we
choose to reduce the length $b$. The modified Bessel function
$K_{\frac{3}{2}} \big(\,\frac{2\pi\,n}{b}\,\sqrt{(\ell_1\,a)^2+
(\ell_2\,c)^2} \,\big)$ goes to zero exponentially as $b\too0$. When
multiplied by $1/b^{3/2}$, the product also goes to zero
exponentially. Therefore the second term in \reff{FPEC} is equal to
zero in the limit $b\too 0$. The result after dimensional reduction
is \beq\begin{split}
 \lim_{b\to 0} \, F_{\!P\!E\!C}&=\dfrac{1}{2\,c}\,\sum_{n=1}^{\infty}\sum_{\ell=1}^{\infty}\dfrac{n}{\ell}
\dfrac{\partial\,}{\partial\,a}K_1\big(\frac{2\,\pi\,n\,\ell\,a}{c}\big)=\dfrac{\pi}{c^2}\,\sum_{n=1}^{\infty}\sum_{\ell=1}^{\infty}n^2\,
K_1^{\,'}\big(\frac{2\,\pi\,n\,\ell\,a}{c}\big)
\end{split}
\eeq{FPEC2} where $K_1^{\,'}\big(x)=dK_1(x)/dx$. The above result
\reff{FPEC2} is exactly equal to the formula [eq.(11)] derived in
\cite{Cavalcanti} for the Casimir force on a piston for a scalar
field obeying Dirichlet boundary conditions in $3D$. We have
therefore shown that the $4D$ PEC piston reduces to the $3D$
Dirichlet piston under dimensional reduction. It would be
interesting to see if high-precision Casimir experiments
\cite{experiments} with real metals in $4D$ can be developed in the
near future to verify this ``reduction" scenario.

We end this section by clarifying a point. As already stated, a $3D$
EM field and a $3D$ scalar field each have one degree of freedom.
For free fields the two are automatically equivalent. However, if we
impose boundary conditions on the EM field, we need to find the
corresponding boundary conditions on the scalar field, which in
general will not be solely  Dirichlet nor Neumann boundary
conditions. Let us look at a concrete example. Consider a $3D$ EM
field confined to an $L_1\times L_2$ rectangular region with PEC
boundary conditions. In radiation gauge ($A_0=0,{\bf\nabla\cdot
A=0}$), PEC conditions yield $A_{||} \eq 0$ and
$\partial_n\,A_{\perp}\eq0$ (where $||$ means parallel to the
surface and $\partial/\partial_n$ denotes the normal derivative).
The mode decomposition is given by
\begin{align}
A_0&=0\nonumber\\
A_1&= a_{n_1n_2} \cos(n_1\pi x_1/L_1)\,\sin(n_2 \pi x_2/L_2)\quad\quad \quad\quad n_1,n_2\ge0\,\,;\,\,\, (n_1,n_2)\ne (0,0).\nonumber\\
A_2&=b_{n_1n_2} \cos(n_2 \pi x_2/L_2)\,\sin(n_1 \pi
x_1/L_1)\nonumber\,.
\end{align}
If $n_1\eq0$ and $n_2\ne0$, then $A_2=0$ and $A_1\ne0$ and if
$n_2\eq0$ and $n_1\ne0$, then $A_1=0$ and $A_2\ne 0$. When $n_1$ and
$n_2$ are both positive, the condition ${\bf\nabla\cdot A=0}$ yields
a relation between the two coefficients: \beq
b_{\,n_1n_2}=-\dfrac{n_1\,L_2}{n_2\,L_1}\,\,a_{\,n_1n_2} \quad \quad
n_1,n_2 \in \mathbb{Z+}\,. \eeq{bn1} The 3D EM field under PEC
conditions is therefore equivalent to a $3D$ scalar field given by
:\beq \phi_{n_1n_2}= a_{n_1n_2} \cos(n_1\pi x_1/L_1)\,\sin(n_2 \pi
x_2/L_2)+ b_{n_1n_2} \cos(n_2 \pi x_2/L_2)\,\sin(n_1 \pi x_1/L_1)
\eeq{phi_n1n2} where $n_1,n_2\ge0, (n_1,n_2)\ne(0,0)$ and
$b_{n_1n_2}$ is related to $a_{n_1n_2}$ via \reff{bn1} when $n_1$
and $n_2$ are both positive. The above scalar field satisfies the
boundary conditions imposed on the EM field but they do not
correspond to either Dirichlet or Neumann boundary conditions on an
$L_1\times L_2$ rectangular region. However, the Casimir energy is
equal to the Neumann Casimir energy since the frequency $\omega$ for
a given mode is given by $(\frac{n_1^2\pi^2}{L_1^2}
+\frac{n_2^2\pi^2}{L_2^2})^{1/2}$ and the sum is over the same modes
$(n_1,n_2)$ (except for the mode $(0,0)$ that appears in the Neumann
case but makes no contribution to the Casimir energy).

\section{No theory recovered under Dirichlet and PEC revisited}

We now consider a $d\p1$ dimensional massless scalar field $\phi$
obeying Dirichlet boundary conditions at the hyperplanes $x^d=0$ and
$x^d=L$. Because the electromagnetic case with PEC and PMC
conditions can be understood as sums over Dirichlet systems
\cite{Ariel-Valery-PMC,Ilana-Thesis}, this case forms a foundation
for understanding more complicated systems. The Fourier
decomposition of the scalar field under Dirichlet conditions is
given by \beq \phi(x^{\mu},x^d)=\sum_{n=1}^{\infty}\phi^n(x^{\mu})
\,\sin(n\,\pi\,x^d/L)\quad \mu=0,1,2,\ldots,d-1\,\,. \eeq{scalar}

After integrating over $x^d$ from $0$ to $L$, we can express the
action for the scalar field as \beq\begin{split} S&=\int
\,\dfrac{1}{2}\,\,\partial_{M}\phi\,\partial^{\,M}\phi \,\,\,d^{\,d+1}x\quad\quad M=0,1,2,...,d\\
&=\dfrac{L}{2}\,\,
\sum_{n=1}^{\infty}\int\Big(\,\dfrac{1}{2}\,\,\partial_{\mu}\phi^n\,\partial^{\,\mu}\phi^n +\dfrac{n^2\,\pi^2}{L^2}\,(\phi^{n})^2\,\Big)\,\,\,d^{\,d}x \,.
\end{split}
\eeq{scalar2} The massless scalar field in $d\p1$ dimensions
decomposes into an infinite tower of massive scalar fields in $d$
dimensions. Note that the sum starts at $n\eq1$. Under dimensional
reduction, i.e. as $L\to 0$, every term becomes infinitely massive
and one does not recover any theory. Correspondingly, the Dirichlet
Casimir piston yields a zero Casimir energy as one dimension is
reduced to zero. The Casimir energy for the Dirichlet Casimir piston
in $d\p1$ dimensions is given by \cite{Ariel-Valery-JHEP}: \beq
E_D=\dfrac{-1}{2\pi}\sum_{\{n_i\}=1}^{\infty}\sum_{\ell=1}^{\infty}
\dfrac{\lambda}{\ell}\,\,K_1(2\,\ell\,\lambda\,a)\quad\word{where}
\lambda=\Bigg[\sum_{i=1}^{d-1}
\dfrac{\pi^2\,n_i^2}{L_i^2}\Bigg]^{1/2}\,.\eeq{ED1} Here $a$ is the
plate separation and the $L_i$ are the lengths of the remaining
$d\m1$ sides. The sum over each $n_i$ starts at $1$. We are
interested in the limit as one of the lengths tends to zero, so
without loss of generality we choose to reduce $L_{d-1}$. As
$L_{d-1}\too 0$, $\lambda\to \infty$ and the modified Bessel
function $K_1(2\,\ell\,\lambda\,a)$ goes to zero exponentially,
yielding a zero Casimir energy: \beq \lim_{L_{d-1}\to0} \,E_D=0\,.
\eeq{DirichLimit} The Dirichlet Casimir energy is zero under
dimensional reduction. We will refer to this as the ``zero
Dirichlet" result.

In the previous section, we showed that the $4D$ PEC piston reduces
to the $3D$ Dirichlet piston under dimensional reduction. We are now
in a position to show this using the ``zero Dirichlet" result. The
$4D$ PEC piston can be decomposed into sums over Dirichlet pistons
of different dimensions \cite{Ariel-Valery-PMC, Ilana-Thesis}: \beq
E_{PEC_{123}}= 2\,E_{D_{123}} + E_{D_{12}}+ E_{D_{13}} +
E_{D_{23}}\,. \eeq{PEC123} Here $E_{PEC_{123}}$ and $E_{D_{123}}$
represent the $4D$ PEC and Dirichlet energies respectively in a
rectangular geometry with three sides of length $L_1,L_2$ and $L_3$,
and $E_{D_{12}}$ represents the $3D$ Dirichlet energy with two sides
of length $L_1$ and $L_2$. We can dimensionally reduce the $4D$ PEC
piston by letting $L_3\to 0$. The ``zero Dirichlet" result implies
that the Dirichlet Casimir energies containing the length $L_3$ go
to zero as $L_3\too 0$. We therefore obtain $\lim_{L_3\to
0}\,E_{PEC_{123}}= E_{D_{12}}$ which is the $3D$ Dirichlet piston
with lengths $L_1$ and $L_2$.

\section{Lower-dimensional system recovered: PMC and Neuman conditions}

Perfect magnetic conductor (PMC) boundary conditions are dual to
perfect electric conductor (PEC) conditions. They are given in any
dimension by $\eta^{\mu}\,F_{\mu\nu}\eq0$, where $\eta^{\mu}$ is a
spacelike vector normal to the hypersurface and $F_{\mu\nu}\equiv
\partial_{\mu} A_{\nu}-\partial_{\nu}A_{\mu}$ is the electromagnetic
field tensor \cite{Ariel-Valery-PMC}. PMC conditions obey the
symmetries of the Maxwell action, namely Lorentz and gauge
invariance. In $3\p1$ dimensions, the conditions at the surface
reduce to ${\bf n\cdot E}=0$ and ${\bf n \times B}=0$, where ${\bf
E}$ and ${\bf B}$ are the electric and magnetic fields and ${\bf n}$
is the vector normal to the surface. Material structures that
approximate PMC's are of current interest because of their
usefulness to communication technologies, in particular as
low-profile antennas \cite{antenna}. An important property of a PMC
is that its surface reflects electromagnetic waves without phase
change of the electric field, in contrast to the $\pi$ phase change
from a PEC \cite{antenna}.

Consider a $d\p1$-dimensional electromagnetic field obeying PMC
conditions on two hyperplanes situated at $x^d\eq0$ and $x^d\eq L$.
The mode decomposition for the gauge fields is given by
\cite{Ariel-Valery-PMC}
 \beq\begin{aligned}
A_{\mu}(x^{\mu},x^d)&\eq \sum_{n=0}^{\infty}A_{\mu}^{(n)}(x^{\mu})
\,\cos\,(n\,\pi\,x^d/L)\\&\eq A_{\mu}^{(0)}(x^{\mu})\p\sum_{n=1}^{\infty}A_{\mu}^{(n)}(x^{\mu})
\,\cos\,(n\,\pi\,x^d/L)\quad\quad\mu=0,1,\ldots,d\m1\\A_{d}(x^{\mu},x^d)&=\sum_{n=1}^{\infty}A_{d}^{(n)}(x^{\mu})
\,\sin\,(n\,\pi\,x^d/L)\,.\end{aligned}\eeq{DecompPMC}

It is convenient to go over to axial gauge $A_d\eq0$. This can be
accomplished with the gauge function \beq\Lambda=
\sum_{n=1}^{\infty}\dfrac{L}{n\,\pi} \,\,A_{d}^{(n)}(x^{\mu})
\,\cos\,(n\,\pi\,x^d/L)\,,\eeq{Lambda2} which does not affect the
form of $A_{\mu}$. In axial gauge, the mode decomposition becomes
 \beq\begin{aligned}
A_{\mu}&\eq A_{\mu}^{(0)}(x^{\mu})\p\sum_{n=1}^{\infty}A_{\mu}^{(n)}(x^{\mu})
\,\cos\,(n\,\pi\,x^d/L)\\A_{d}&=0\,.\end{aligned}\eeq{DecompPMC2}

We now substitute the mode decomposition \reff{DecompPMC2} into the
$d\p1$-dimensional Maxwell action \reff{action}. After integrating
over $x^d$ from $0$ to $L$, the resulting action is:
\beq\begin{aligned} S&= \dfrac{L}{2}\int -\dfrac{1}{4}
\,F_{\mu\nu}^{(0)}F^{{\mu\nu}^{(0)}}d^{\,d}x \,\,+\,\,
\dfrac{L}{2}\int \sum_{n=1}^{\infty}\Big\{-\dfrac{1}{4}
\,F_{\mu\nu}^{(n)}F^{{\mu\nu}^{(n)}} + \dfrac{1}{2}
\dfrac{n^2\pi^2}{L^2}\,A_{\mu}^{(n)}A^{{\mu}^{(n)}}\Big\}\,d^{\,d}x
\end{aligned}\eeq{action3}
where $F_{\mu\nu}^{(0)}\equiv \partial_{\mu}
A_{\nu}^{(0)}-\partial_{\nu}A_{\mu}^{(0)}$. Under PMC conditions,
the Maxwell action in $d\p1$ dimensions decomposes into two sectors:
Maxwell in $d$ dimensions plus an infinite tower of $d$-dimensional
massive gauge fields. Under dimensional reduction, where we let
$L\to0$, the mass of the gauge fields becomes infinite so that the
massive sector can be ignored. Therefore, for any starting
dimension, we recover the lower-dimensional version of the original
system: the $d$-dimensional Maxwell action. In contrast, for PEC
conditions, the lower-dimensional version was recovered only in $4D$
and for Dirichlet boundary conditions no theory is recovered at all.

We now show that the PMC Casimir piston in $d\p1$ dimensions reduces
to the $d$-dimensional PMC piston under dimensional reduction. For
this purpose, a convenient formula for the Casimir force on a PMC
piston in $d\p1$ dimensions is expression $A.6$ found in
\cite{Ariel-Valery-PMC}: \beq F_{\!P\!M\!C}=
-\sum_{p=1}^{d-1}\,\,\sum_{q=0}^{d-p-1}
\dfrac{\pi}{2^{d-q+1}}\,(d\!-\!1\!-\!2p\!-\!q)\,\,\xi^{\,d-q-1}_{\,1,k_2,k_3,..,
k_p}\,\dfrac{a_{k_2-1}\ldots a_{k_p-1}}
{(a_{d-q-1})^{p+1}}\dfrac{\partial\,}{\partial\,a}
\big\{a\,R_{p}\big\}\eeq{FPMCalt} where  \beq R_{p}
=\sum_{n=1}^{\infty}\,\sum_{\ell_1=1}^{\infty}\sum_{\substack{\ell_i=-\infty\\i=2,\ldots,
p}}^{\infty}\dfrac{4\,\,n^{\frac{p+1}{2}}}{\pi}\,\dfrac{\,K_{\frac{p+1}{2}}
\big(\,2\pi\,n\,\sqrt{(\ell_1\frac{a}{a_{d-q-1}})^2+\cdots+(\ell_p\,\frac{a_{k_p-1}}{a_{d-q-1}})^2}\,\,
\,\big)}{\left[(\ell_1\frac{a}{a_{d-q-1}})^2+\cdots+(\ell_p\frac{a_{k_p-1}}{a_{d-q-1}})^2\right]^{\tfrac{p+1}
{4}}}\,. \eeq{RIpalt} In the above formula, $a$ is the plate
separation and the lengths of the other $d\m 1$ sides are
$a_1,a_2,\ldots,a_{d-1}$. We are interested in evaluating the limit
of $F_{\!P\!M\!C}$ as one of the lengths is reduced to zero. Without
loss of generality we choose the length to be $a_{d-1}$.  This
length appears in \reff{FPMCalt} when $q\eq0$ in $a_{d-q-1}$, i.e.
it appears in the denominator in \reff{FPMCalt} and in the argument
of the modified Bessel function of $R_{p}$ in \reff{RIpalt} when
$q\eq0$. We are therefore interested in taking the limit of only the
$q\eq0$ terms. In the limit $a_{d-1}\too 0$ it is easy to see that
those terms are zero, because the modified Bessel functions that
appear in $R_{p}$ go to zero exponentially in this limit. We
obtain\footnote{The ordered symbol $\xi^{\,d-q-1}_{\,1,k_2,k_3,..,
k_p}$ introduced in \cite{Edery} and the product $a_{k_2-1}\ldots
a_{k_p-1}$ do not contain $a_{d-1}$.} $\lim_{a_{d-1}\to 0}
\,(a_{d-1})^{-p-1}\,\partial\,(a\,R_{p})/\partial\,a=0$. After
taking the limit, the sum over $p$ now runs from $1$ to $d\m2$
instead of $d\m1$. The sum over $q$ runs from $1$ to $d-p-1$. This
is equivalent to $q$ running from $0$ to $d\m p\m 2$ if we replace
$d$ by $d\m1$ in the summand of \reff{FPMCalt}. The final result is
thus an identical formula to $F_{PMC}$, except that $d$ is now
replaced by $d\m 1$: \beq \lim_{a_{d-1}\to0}F_{\!P\!M\!C}=
-\sum_{p=1}^{d-2}\,\,\sum_{q=0}^{d-p-2}
\dfrac{\pi}{2^{d-q}}\,(d\!-\!2\!-\!2p\!-\!q)\,\,\xi^{\,d-q-2}_{\,1,k_2,k_3,..,
k_p}\,\dfrac{a_{k_2-1}\ldots a_{k_p-1}}
{(a_{d-q-2})^{p+1}}\dfrac{\partial\,}{\partial\,a}
\big\{a\,R_{p}\big\}. \eeq{FPMCalt2} where $R_p$ is given by
\reff{RIpalt} with $d$ replaced by $d\m1$. We therefore recover the
lower-dimensional version of the PMC Casimir force under dimensional
reduction.

We now consider a $d\p1$ dimensional massless scalar field $\phi$
obeying Neumann boundary conditions $\eta^{\mu}\partial_{\mu}\phi=0$
at the hyperplanes $x^d=0$ and $x^d=L$, where $\eta^{\mu}$ is the
spacelike vector in the $x^d$ direction normal to the hyperplanes.
These conditions are Lorentz invariant. The Fourier decomposition of
the scalar field under Neumann conditions is \beq\begin{split}
\phi(x^{\mu},x^d)&=\sum_{n=0}^{\infty}\phi^n(x^{\mu})
\,\cos(n\,\pi\,x^d/L)\quad\quad \mu=0,1,2,\ldots,d-1\\&=
\phi^{(0)}(x^{\mu}) +\sum_{n=1}^{\infty}\phi^n(x^{\mu})
\,\cos(n\,\pi\,x^d/L)\,.
\end{split} \eeq{scalar3}

After integrating over $x^d$ from $0$ to $L$ the action for the scalar field can be expressed as
\beq\begin{split}
S&=\int
\,\dfrac{1}{2}\,\,\partial_{M}\phi\,\partial^{\,M}\phi \,\,\,d^{\,d+1}x\quad\quad M=0,1,2,...,d\\
&=\dfrac{L}{2}\,\,\int\dfrac{1}{2}\,\,\partial_{\mu}\phi^{(0)}\,\partial^{\,\mu}\phi^{(0)} \, d^{\,d}x  +\dfrac{L}{2}\,\,
\sum_{n=1}^{\infty}\int\Big(\,\dfrac{1}{2}\,\,\partial_{\mu}\phi^n\,\partial^{\,\mu}\phi^n +\dfrac{n^2\,\pi^2}{L^2}\,(\phi^{n})^2\,\Big)\,\,\,d^{\,d}x \,.
\end{split}
\eeq{scalar2a} The massless scalar field in $d\p1$ dimensions under
Neumann conditions has decomposed into a massless sector with a
scalar field in $d$ dimensions plus a massive sector with an
infinite tower of $d$-dimensional massive scalar fields $\phi^{(n)}$
of mass $m_n=n\,\pi/L$.  Under dimensional reduction, $L\to 0$, the
masses go to infinity, so that the massive sector can be ignored. We
therefore recover the lower-dimensional version of the original
system: a $d$-dimensional massless scalar field.

We have encountered two special systems, the PMC and Neumann
systems, where the lower-dimensional version of the original system
in any dimension is recovered under dimensional reduction. What do
these boundary conditions have in common? They both are
``variational'' conditions that can be obtained by minimizing the
action with the field free to vary on the surface. Such conditions
arise naturally in bag models \cite{Jaffe}.

We vary the action $S\,[\Phi]$ with respect to the field $\Phi$,
where $S\,[\Phi]$ vanishes outside a bounded region. Assuming the
equations of motion are satisfied, we obtain the following boundary
term (which must be set to zero) \beq \int \,
\,\partial_{\mu}\,\Big(\,\dfrac{\partial\mathscr{L}}{\partial(\partial_{\mu}\Phi)}\,\delta\Phi\Big)\,d^{d+1}x=
\int \eta_{\mu}\dfrac{\partial
\mathscr{L}}{\partial(\partial_{\mu}\Phi)}\,\delta\Phi\,\,d\,\sigma=0\eeq{ghf}
where $\eta^{\mu}$ is a spacelike vector normal to the timelike
hypersurface $\sigma$. If $\Phi$ is allowed to vary on the
boundary\footnote{ Dirichlet and PEC boundary conditions do not
fulfill this criteria. For the case of Dirichlet, the field $\phi$
is zero on the boundary and for the PEC case, the gauge fields
$A_{\nu}$ are zero on the boundary except for one component.} i.e.
$\delta \Phi\ne 0$, we then obtain the following Lorentz invariant
boundary condition: \beq \eta_{\mu}\,\dfrac{\partial
\mathscr{L}}{\partial(\partial_{\mu}\Phi)}=0\,. \eeq{noflux} If
$\Phi$ is a Klein-Gordon scalar field $\phi$, then
$\mathscr{L}=\frac{1}{2}\,\,\partial_{\mu}\phi\,\partial^{\,\mu}\phi$
and we obtain Neumann boundary conditions $\eta^{\mu}\partial_{\mu}
\phi=0$, and if $\Phi$ is an abelian gauge field $A_{\nu}$, then
$\mathscr{L}=-\frac{1}{4} \,F_{\mu\nu}F^{\mu\nu}$ and we obtain PMC
conditions $\eta^{\mu}F_{\mu\nu}=0$. In both cases, there is no
momentum flux through the hypersurface $\sigma$ even though the
field is free to vary on the surface. For the scalar field,
$\eta_{\mu}\,T^{\mu0}=(\eta_{\mu}\,\partial^{\mu}\phi)\,\partial^{0}\phi$,
and this is equal to zero for Neumann boundary conditions. For an EM
field, $\eta_{\mu}\,T^{\mu0}=
(\eta_{\mu}\,F^{\mu\alpha})\,F^{0}_{\;\;\alpha}$, and this is equal
to zero for PMC conditions.

For PMC or Neumann boundary conditions, where the fields are free to
vary on the surface, the fields have zero modes. These zero modes
survive after dimensional reduction and one recovers the
lower-dimensional version of the original action. This would also
occur in chiral models, since the Dirac field would not be fixed on
the boundary and would have zero modes. This is in contrast to
Dirichlet and PEC conditions. For Dirichlet, the scalar field has no
zero modes and under dimensional reduction it becomes infinitely
massive. PEC conditions are more interesting. One does not recover
the lower-dimensional Maxwell action under dimensional reduction but
the action of a scalar field. Only one component of the gauge fields
has a zero mode.

\section{Conclusion}

We have studied dimensional reduction for abelian gauge fields under
PMC and PEC conditions and for scalar fields under Dirichlet and
Neumann boundary conditions. Our investigation was not restricted to
extra dimensions and included dimensional reduction of $4D$ systems.
In particular, PEC and PMC conditions can be viewed as idealized
material boundary conditions for EM fields in 4D. We showed that for
PEC conditions, a $D$-dimensional EM field reduces under dimensional
reduction to a $D\m1$ scalar field, and not its lower-dimensional
version, a $D\m1$ EM field. In particular, we showed that the $3D$
Dirichlet piston can be obtained via dimensional reduction from a
more realistic $4D$ EM piston obeying perfect conductor conditions.
While the $3D$ scalar field system was a toy model and inaccessible
experimentally, there is now the possibility that high-precision
Casimir experiments \cite{experiments} involving real metals in $4D$
could be developed to test this scenario or something similar to it.
We noted that a $3D$ EM field and a $3D$ scalar field both have one
degree of freedom, but are equivalent only as long as the scalar
field takes into account the boundary conditions imposed on the EM
field. In particular, a $3D$ EM field confined to a rectangular
geometry under PEC conditions is equivalent to the $3D$ scalar field
given by \reff{phi_n1n2}, which does not correspond to Dirichlet or
Neumann boundary conditions even though the mode frequencies are
equal to those of Neumann boundary conditions.

For Dirichlet boundary conditions, we found that under dimensional
reduction no theory is recovered and the Casimir force on a
Dirichlet piston is zero. This ``zero Dirichlet" result is
particularly useful in showing that the $4D$ PEC Casimir piston
reduces to the $3D$ Dirichlet Casimir piston under dimensional
reduction. We identified two special boundary conditions, perfect
magnetic conductor (PMC) conditions for EM fields and Neumann
conditions for scalar fields, where dimensional reduction yields the
lower-dimensional version of the action in any dimension and
verified explicitly this result for the PMC Casimir piston. These
two cases represent ``variational'' boundary conditions where the
action vanishes outside a bounded region and the field is not fixed
on the surface. As a consequence, PMC and Neumann conditions have
zero modes and these yield the lower-dimensional version of the
action under dimensional reduction.

\begin{appendix}
\def\theequation{A.\arabic{equation}}
\setcounter{equation}{0}
\section{Action for PEC system using axial gauge $A_d\eq0$}

In this appendix, using axial gauge $A_d\eq0$, we derive the action
\reff{action2} starting with the mode decomposition \reff{Decomp}
for the EM field under PEC conditions. For axial gauge $A_d\eq0$,
the gauge function \reff{Lambda} has to be modified to\beq\Lambda=
\sum_{n=1}^{\infty}-\dfrac{L}{n\,\pi} \,\,A_{d}^{(n)}(x^{\mu})
\,\sin\,(n\,\pi\,x^d/L)- A_d^{(0)}(x^{\mu})\,x^d\,.\eeq{Lambda22} A
gauge transformation with the above gauge function yields the
following mode decomposition for the gauge fields
\beq\begin{aligned} A_{\mu}(x^{\mu},x^d)&=
\sum_{n=1}^{\infty}A_{\mu}^{(n)}(x^{\mu})
\,\sin\,(n\,\pi\,x^d/L)-\partial_{\mu}\phi\,x^d\\A_{d}&=0\end{aligned}\eeq{Amu2}
where the first term in \reff{Lambda22} has been absorbed into a
redefinition of $A_{\mu}^{(n)}(x^{\mu})$ and $\phi(x^\mu)\equiv
A_d^{(0)}(x^{\mu})$ represents a scalar field. We need to evaluate
the Maxwell action \reff{action} with the mode decomposition
\reff{Amu2}. The tensor $F_{\mu\nu}$ is given by \beq
F_{\mu\nu}\equiv\partial_{\mu}A_{\nu}-\partial_{\nu}A_{\mu}=\sum_{n=1}^{\infty}\left(\partial_{\mu}A_{\nu}^{(n)}-
\partial_{\nu}A_{\mu}^{(n)}\right)\sin(n\,\pi x^d/L)=\sum_{n=1}^{\infty}F_{\mu\nu}^{(n)}\sin(n\,\pi x^d/L)\,.\eeq{fdsgh}
The scalar field $\phi$ does not appear in $F_{\mu\nu}$ because the
partial derivatives commute i.e. \linebreak
$(-\partial_{\mu}\partial_{\nu}+\partial_{\nu}\partial_{\mu})\,\phi=0$.
Squaring the tensor and integrating over the $d\p1$-dimensional
spacetime yields \beq\int \,F_{\mu\nu}\,F^{\mu\nu} \,d^{\,d+1}x =
\dfrac{L}{2}\int \sum_{n=1}^{\infty}
\,F_{\mu\nu}^{(n)}F^{{\mu\nu}^{(n)}}\,d^{\,d}x \eeq{FF1} where we
have integrated over $x^d$ using the orthogonal property \beq
\int_0^L \sin(n\,\pi x^d/L)\,\sin(m\,\pi x^d/L)\,dx^d =
\dfrac{L}{2}\,\delta_{nm}\,\quad n,m\ge 1.\eeq{sines} The tensor
$F_{\mu\,d}$ is given by \beq
F_{\mu\,d}\equiv\partial_{\mu}A_d-\partial_d\,A_{\mu}=-\partial_d\,A_{\mu}=\sum_{n=1}^{\infty}-\dfrac{n\pi}{L}A_{\mu}^{(n)}\cos(n\,\pi
x^d/L)+\partial_{\mu}\phi \,.\eeq{Fmud22} Again, squaring the tensor
and integrating yields \beq \int \,F_{\mu\,d}\,F^{\mu\,d}
\,d^{\,d+1}x= -\dfrac{L}{2}\int \sum_{n=1}^{\infty}
\left(\dfrac{n\pi}{L}\right)^2 A_{\mu}^{(n)}A^{{\mu}^{(n)}} d^dx
-L\int
\partial_{\mu}\phi\,\partial^{\mu}\phi \,d^{\,d}x \eeq{FF2}
where $F^{\mu\,d}=-\partial^d\,A^{\mu}=\partial_d\,A^{\mu}$ and we
have integrated over $x^d$ using the following results:\beq \int_0^L
\cos(n\,\pi x^d/L)\,\cos(m\,\pi x^d/L)\,dx^d =
\dfrac{L}{2}\,\delta_{nm}\quad ;\quad\int_0^L \cos(n\,\pi
x^d/L)\,dx^d =0\,\quad n,m\ge1.\eeq{cosines} With \reff{FF1} and
\reff{FF2}, the $d\p1$-dimensional Maxwell action \reff{action} can
be expressed as \beq\begin{aligned}S&=\int -\dfrac{1}{4}
\,F_{MN}F^{MN}
\,d^{\,d+1}x \fivequad \qquad M,N=0,1,\ldots,d\\
&=\int -\dfrac{1}{4} \,F_{\mu\nu}F^{\mu\nu} \,d^{\,d+1}x +\int
-\dfrac{1}{2} \,F_{\mu\,d}F^{\mu\,d} \,d^{\,d+1}x \qquad \quad
\mu,\nu=0,1,\ldots,d-1\\&=\dfrac{L}{2}\int
\,\partial_{\mu}\phi\,\partial^{\,\mu}\phi \,\,d^{\,d}x+
\dfrac{L}{2}\int \sum_{n=1}^{\infty}\left\{-\dfrac{1}{4}
\,F_{\mu\nu}^{(n)}F^{{\mu\nu}^{(n)}} + \dfrac{1}{2}
\dfrac{n^2\pi^2}{L^2}\,A_{\mu}^{(n)}A^{{\mu}^{(n)}}\right\}d^{\,d}x
\end{aligned}\eeq{final action}
which is equal to the action \reff{action2}.
\end{appendix}

\section*{Acknowledgments}
A.E. acknowledges support from a discovery grant of the National
Science and Engineering Research Council of Canada (NSERC). N. G.
was supported in part by the National Science Foundation (NSF)
through grant PHY05-55338, and by Middlebury College. A.E. and N.G.
would like to thank the Kavli Institute for Theoretical Physics
(KITP) for their hospitality where portions of this work were
completed. This research was  supported in part by the National
Science Foundation under Grant No. PHY05-51164.

\end{document}